\documentclass{nature_mod}

\usepackage{diagbox}
\usepackage{url}
\usepackage{caption}
\usepackage{epsfig}
\usepackage{graphicx}
\usepackage{xcolor}
\usepackage{amssymb,amsmath,caption}
\usepackage{lineno}
\usepackage{aas_macros}
\usepackage{threeparttable}
\usepackage{booktabs}
\usepackage{longtable}
\usepackage{multibib}
\usepackage{float}
\usepackage{hyperref,cleveref}
\newcites{methods}{Methods References}
\definecolor{darkblue}{RGB}{0,0,128}

\long\def\symbolfootnote[#1]#2{\begingroup%
\def\thefootnote{\fnsymbol{footnote}}\footnote[#1]{#2}\endgroup}

\title{Hardware-aware quantum attention for fast radio burst identification}

\author{
Li He$^{1}$\thanks{These authors contributed equally to this work.},
Xuan Yang$^{2\,*}$,
Feng Xiong$^{1\,*}$,
Songbo Zhang$^{2,\dagger}$\thanks{Email: sbzhang@pmo.ac.cn},
Qiuhao Chen$^{1,\dagger}$\thanks{Email: chenqiuhao@tgqs.net},
Yuchen He$^{1}$,
Yunxiang Yang$^{1}$,
Jun Lu$^{1}$,
Claudio Furtado$^{3,4}$,
Amilcar R. Queiroz$^{5,4}$\thanks{Email: amilcarq@df.ufcg.edu.br},
Xinping Deng$^{6,7}$\thanks{Email: xinping.deng@foxmail.com},
Yang Wu$^{6,7}$,
Xuefeng Wu$^{2,8}$\thanks{Email: xfwu@pmo.ac.cn}
}

\begin{document}
\maketitle

\begin{affiliations}
\item Yangtze Delta Industrial Innovation Center of Quantum Science and Technology, Suzhou 215000, China
\item Purple Mountain Observatory, Chinese Academy of Sciences, Nanjing 210023, China
\item Departamento de Física, Universidade Federal da Paraíba, 58051-970, João Pessoa, PB, Brazil 
\item CIQUANTA-PB (Centro Internacional de Computação Quântica), Estação Cabo Branco, Avenida João Cirillo da Silva,  58046-010, João Pessoa, PB, Brazil
\item Unidade Acadêmica de Física, Universidade Federal de Campina Grande, R. Aprígio Veloso 882, Bairro Universitário, 58429-900, Campina Grande, PB, Brazil
\item The 54th Institute of China Electronics Technology Group Corporation, 498 Zhongshan W Rd, Shijiazhuang 050002, Hebei, China
\item China-Brazil Belt and Road Joint Laboratory on Radio Astronomy Technology, 498 Zhongshan W Rd, Shijiazhuang 050002, Hebei, China
\item School of Astronomy and Space Sciences, University of Science and Technology of China, Hefei 230026, China
\end{affiliations}

\bigskip

\begin{abstract}
Fast radio bursts (FRBs) are millisecond-duration extragalactic pulses whose discovery requires searches over large signal-parameter spaces and the rejection of candidate sets dominated by radio-frequency interference and noise. Integrating quantum processors into FRB searches requires a workflow connecting telescope data to hardware-compatible models. Here we develop a pipeline that converts raw search-mode PSRFITS data into dynamic-spectrum segments and identifies FRB-like signals using a hybrid Quantum Vision Transformer (QViT). On labelled FAST data, the selected QViT achieves a mean accuracy of 94.00\% and recall of 98.30\%, with similar performance to a compact classical Vision Transformer. Recall reaches 97.11\% on an FRB source excluded from training and model selection. Real-device execution on 82 segments yields a mean simulator--hardware Hellinger fidelity of $0.9161\pm0.0074$. The workflow connects raw telescope data to quantum-assisted identification and provides an experimental basis for assessing the measurement requirements that must be addressed before survey-scale deployment.
\end{abstract}

\clearpage


Fast radio bursts (FRBs) are bright, millisecond-duration radio transients of extragalactic origin~\cite{Lorimer07,Thornton13,Spitler16}. Their detection usually involves searching over dispersion measure, pulse width and arrival time, followed by the rejection of candidates dominated by radio-frequency interference (RFI) and noise~\cite{Cordes03,Petroff19AARv,Zhang20_PTD1,Li21,Niu21,Xu22,Li26}. Machine-learning classifiers have reduced this inspection burden~\cite{Fetch,Yang21}, and recent approaches identify bursts directly in non-dedispersed dynamic spectra~\cite{Chen26}. 
A model that examines segments formed directly from telescope observations can identify FRB-like signals without requiring candidate-derived alignment.

FRB identification is already tractable with classical artificial intelligence. Against this background, we investigate quantum machine learning~\cite{cerezo2021,biamonte_quantum_2017,du_gentle_2025,Kordzanganeh21_QML_RA} as a complementary and exploratory route, rather than as a replacement for established classical approaches at this stage. Parameterized quantum circuits provide trainable feature maps with structures that differ from those of classical neural-network layers~\cite{schuld2019qmlFeatureHilbertSpace,huang_power_2021,yu_non-asymptotic_2024}. Quantum-native subroutines may eventually offer different resource scaling for selected transformations or searches~\cite{grover1996,harrow2009,gilyen2019,montanaro2016}, but any practical benefit must be assessed together with data loading and state preparation, measurement and readout, fault-tolerance overheads, and classical--quantum data movement~\cite{aaronson2015fineprint,cerezo_challenges_2022,fowler2012surface,hoefler2023disentangling,zhang_bits_2026}.

Attention-based architectures are well suited to learning relationships among distributed features in a time--frequency representation~\cite{dosovitskiy2020}. In a hybrid Quantum Vision Transformer (QViT), parameterized quantum circuits replace the learned linear projections that generate query, key and value representations, after which conventional scaled dot-product attention performs token-to-token mixing~\cite{Cherrat24_QViT,Zhang25_HQViT}. Present quantum processors constrain the number of qubits, circuit depth, measurement budget and executable model size. We therefore co-design the attention dimensions and quantum circuits for real-device execution and compare them with a classical Vision Transformer in the same low-dimensional regime. This approach uses current hardware to explore a complementary algorithmic route while making its practical constraints explicit.

Here we develop a hardware-aware QViT for identifying FRB-like signals in dynamic-spectrum segments constructed from raw search-mode PSRFITS data~\cite{Hotan04}. The identification workflow links data preparation, image construction and hybrid quantum--classical inference. Using labelled FAST observations~\cite{Jiang20}~\cite{Jiang20} of two repeating FRBs~\cite{Lorimer07} and one rotating radio transient (RRAT)~\cite{McLaughlin06}, we compare QViT with a classical ViT, evaluate transfer to an FRB source excluded from training and model selection, and test a hardware-compatible configuration on a real quantum processor.

\section*{Results}
\label{sec:results}
\begin{figure*}
\centering
\includegraphics[width=0.8\linewidth]{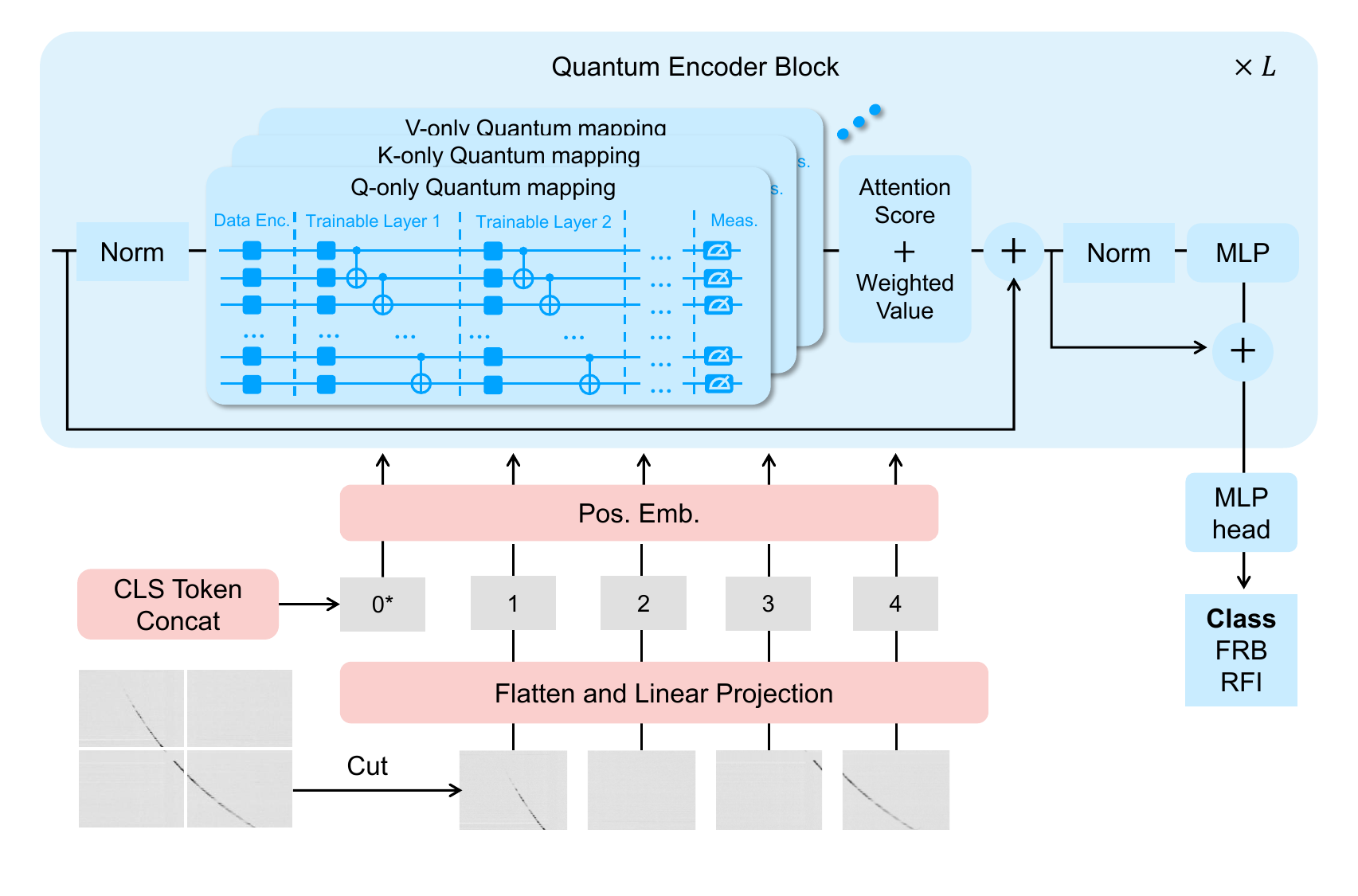}
\caption{Architecture of the QViT identification model. Dynamic-spectrum images constructed from PSRFITS data are resized to $288\times288$ pixels, converted into tokens and passed to the hybrid transformer. Parameterized quantum circuits generate the query, key and value representations, while attention mixing and classification remain classical. The model returns an FRB-like probability for each input segment.}
\label{fig:qvit_flowchart}
\end{figure*}

The identification workflow constructs dynamic-spectrum images from selected intervals of PSRFITS data and assigns an FRB-like probability to each segment (Figure~\ref{fig:qvit_flowchart}). We evaluate the models on labelled FAST segments, followed by a source-held-out test and a comparison of simulator and real-device outputs.
With the default preprocessing, blocks of 16 adjacent frequency channels and 16 consecutive time samples are averaged before each image is resized to $288\times288$ pixels. Conventional single-pulse searches provide labels and identify the intervals used to construct positive examples; their candidate scores are not supplied to the model. This defines the inference path from raw telescope data to segment-level FRB identification.

The training sources comprise two repeating FRBs, FRB 20220529~\cite{Li26,Yang26} (1243 pulses) and FRB 20201124A~\cite{Xu22} (1093 pulses), and the rotating radio transient RRAT J1913+1330~\cite{Zhang23_1913} (664 pulses). These 3000 astrophysical pulses form the FRB-like class, and 3000 segments from single-pulse candidates that manual vetting rejected as RFI or other non-astrophysical signals form the negative class. The 3000 positive and 3000 negative segments are divided into training ($2400+2400$), validation ($300+300$) and test ($300+300$) partitions, and a separate source-held-out set of $114+114$ segments from FRB 20220912A~\cite{Zhang23_0912A} forms an additional test set that is excluded from training and model selection and is used to validate both models on a source absent from training (Extended Data Table~\ref{tab:datasets}). Within each model family, candidate configurations are ranked on the validation partition; the test partition provides an independent evaluation of the selected models, and FRB 20220912A tests transfer to an unseen source. All reported evaluations use labelled segments drawn from FAST observations.

\subsubsection*{Comparison with classical attention}

ViT and QViT were evaluated over the same predefined ranges for corresponding architectural parameters. Within each model family, we selected the configuration with the highest mean recall on the validation set across 10 independent seeds. The selected configurations were then evaluated on the test set, which was not used for model selection. Table~\ref{tab:main_performance} reports the selected configurations as the mean and standard deviation over 10 independent seeds.

On the main test set, QViT achieves an accuracy of $94.00\pm0.48\%$ and a recall of $98.30\pm0.60\%$, compared with $93.83\pm0.60\%$ and $98.23\pm0.22\%$, respectively, for ViT. QViT attains a precision of $90.52\pm0.62\%$ and an $F_2$ score of $96.64\pm0.50\%$; the corresponding ViT values are $90.30\pm1.02\%$ and $96.54\pm0.26\%$. The differences in mean accuracy, recall, precision and $F_2$ are 0.17, 0.07, 0.22 and 0.10 percentage points, respectively. The selected configurations therefore show similar observed performance on this dataset, with recall above 98\% and precision close to 90\%. Because configurations were ranked on the validation partition and the decision threshold was fixed there, the test partition provides an independent evaluation of the selected models.

We examined the dependence of recall on the signal-to-noise ratio (SNR) reported by the single-pulse search (see Methods for details), as weak signals are generally more challenging cases for transient identification~\cite{Yang21}. The provisional analysis divides the test data into low-SNR ($\mathrm{SNR}<12$) and high-SNR ($\mathrm{SNR}\geq12$) subsets (Extended Data  Table~\ref{tab:snr} and Extended Data Fig.~\ref{fig_recall_snr_compare}). QViT attains a low-SNR recall of $96.62\pm1.39\%$, compared with $95.88\pm0.93\%$ for ViT. The corresponding high-SNR recalls are $98.79\pm0.53\%$ and $98.92\pm0.23\%$, respectively.

\subsubsection*{Generalization to a source excluded from training}

The selected models were evaluated on FRB 20220912A using the trained weights and thresholds fixed with validation data. On this source-held-out dataset, QViT reaches an accuracy of $94.91\pm0.72\%$ and a recall of $97.11\pm1.31\%$, compared with $95.09\pm0.54\%$ and $97.90\pm0.74\%$, respectively, for ViT (Table~\ref{tab:main_performance}). QViT has a precision of $93.07\pm1.81\%$ and an $F_2$ score of $96.26\pm0.82\%$, whereas ViT achieves $92.69\pm0.51\%$ and $96.80\pm0.64\%$. Both models retain high recall for a source absent from training. QViT has a slightly higher mean precision, while ViT has higher mean accuracy, recall and $F_2$ on this dataset.

\subsubsection*{Consistency between simulated and real-device outputs}

The trained quantum-attention circuits were executed on the Baihua quantum processor in a hybrid quantum--classical configuration. Only the Q/K/V projection circuits were submitted to the quantum processing unit (QPU). PSRFITS reading, frequency--time averaging and dynamic-spectrum image construction were performed on the CPU, whereas image preprocessing, tokenization, attention weighting, normalization, feedforward processing and classification remained on conventional GPU-accelerated hardware. Real-device measurements were then passed through the remaining classical layers to assess their downstream effect on segment classification. Because misidentification is most likely for weak signals, and real-device execution is resource-limited, the hardware-validation subset was drawn from the low-SNR portion of the main test partition. It contains 82 labelled segments, comprising 41 low-SNR FRB-like and 41 RFI segments. It uses a six-qubit configuration, 204 Q/K/V circuit evaluations per image and 20,000 shots per circuit (Fig.~\ref{fig_real_qc_used}), corresponding to $4.08\times10^6$ shots per image. This is a hardware-compatible configuration of the same QViT projection module; its register size, circuit depth and token count are set by what the device can execute rather than by the architecture sweep.

\begin{figure*}
\centering
\includegraphics[width=\linewidth]{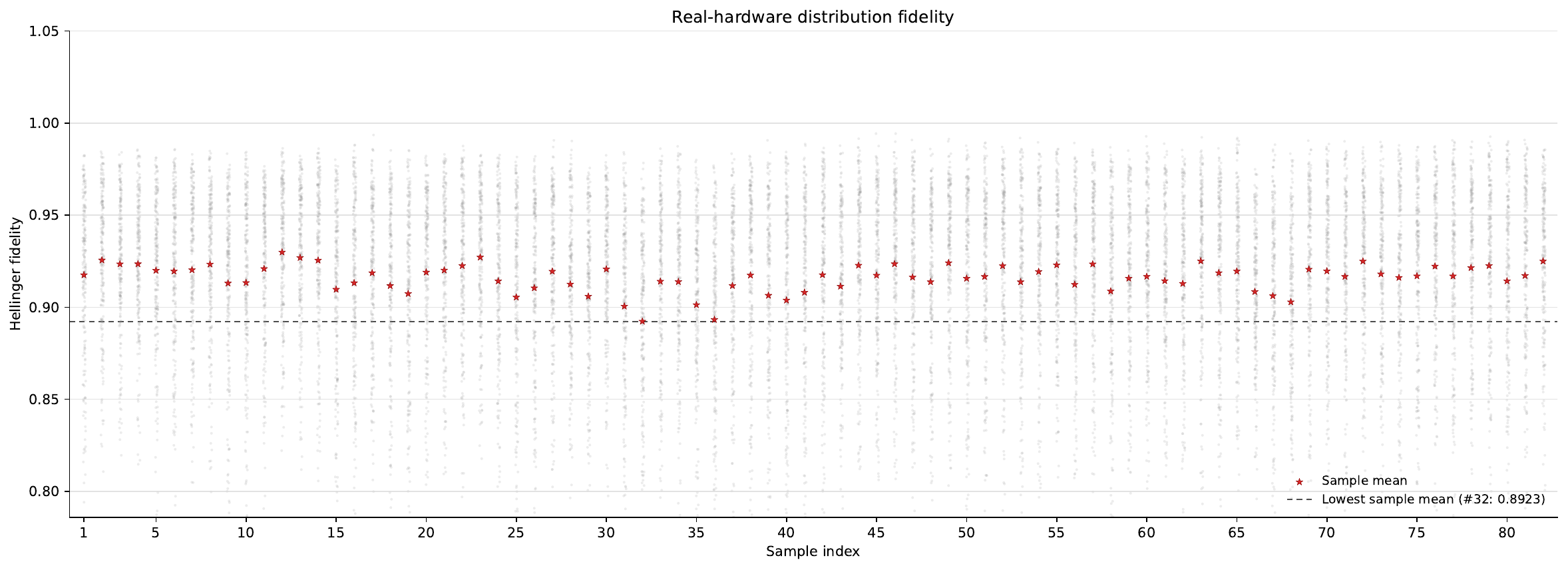}
\caption{Hellinger fidelity between simulated and real-device measurement distributions for the real-hardware validation samples. For each segment, the fidelity is averaged over its 204 Q/K/V circuit evaluations. Across 82 hardware-validation segments, each circuit was executed with 20,000 shots, and the minimum segment-level mean fidelity is 0.8923.}
\label{fig_real_hellinger}
\end{figure*}

We compared the real-device outputs with the corresponding simulator outputs at two levels: the measurement probability distributions and the Pauli-$Z$ expectation-value vectors used as quantum features. The mean Hellinger fidelity between the simulated and measured distributions is $0.9161\pm0.0074$ (Fig.~\ref{fig_real_hellinger}). The mean cosine similarity between the simulated and measured feature vectors is $0.9853\pm0.0022$ (Extended Data Fig.~\ref{fig_real_cosine}). These measurements indicate agreement in the output distributions and in the direction of the feature vectors for the tested inputs, while allowing deviations in the final classification decisions.

Table~\ref{tab:real_confusion} provides a complementary comparison of the downstream classification outcomes for these same 82 images. Of the 41 FRB-like segments, 33 are correctly identified using simulated projections and 34 using real-device projections. Of the 41 RFI segments, 40 are correctly identified with simulated projections and 34 with real-device projections.

\section*{Discussion}
\label{sec:discussion}

This work connects raw telescope observations to quantum-assisted FRB identification within a single processing workflow. Designing the QViT around executable circuits makes the quantum projection module part of that workflow and allows its behaviour to be tested on a real processor. The resulting implementation provides a starting point for evaluating successive quantum models against a common observational task as hardware capabilities develop.
The quantum circuits replace only the Q/K/V projections; attention mixing and the downstream transformer and classification layers remain classical. Present limits on qubit number, circuit depth, token sequence and measurement throughput determine the compact model used here.

The real-device results provide a component-level consistency test. The simulator and hardware produce broadly similar projection outputs, but the class-wise outcomes in Table~\ref{tab:real_confusion} show that projection-level agreement does not guarantee identical final decisions. Given the limited validation set, these results are a component-level consistency check rather than an accuracy or recall benchmark for the device. Repeated circuit evaluations, finite-shot measurement and classical--quantum communication remain the principal obstacles to scaling the present implementation.

Larger qubit registers and more reliable circuits could support wider attention representations and deeper trainable projections within the same workflow. Faster readout, efficient batching and reduced communication overhead would also be needed to translate these improvements into higher processing throughput. If model capacity and end-to-end speed improve together, quantum-assisted identification could become relevant to data-intensive projects such as SKA~\cite{Dewdney09_SKA}, FASTCA~\cite{Jiang24_FASTCA} and to high-throughput coherent all-sky monitors such as CASM-256~\cite{Connor26_CASM}. The present implementation provides a framework in which those developments can be tested against the requirements of astronomical searches.

\section*{Methods}
\subsubsection*{Observational data and construction of labels}\label{subsubsec:datasets}

The analysis uses raw search-mode intensity data stored in PSRFITS format from FAST observations~\cite{Jiang20} of two repeating FRBs (FRB 20220529~\cite{Li26, Yang26} and FRB 20201124A~\cite{Xu22}) and one rotating radio transient (RRAT J1913+1330~\cite{Zhang23_1913}). A separate FRB source, FRB 20220912A~\cite{Zhang23_0912A}, is excluded from training and model selection and is used only for source-held-out evaluation. Here, ``raw'' denotes the channelized search-mode intensity array read from the PSRFITS file.

The datasets were obtained with the FAST 19-beam L-band receiver over 1000--1500\,MHz, split into 4096 frequency channels, digitized at 8-bit sampling with a ROACH2 backend and recorded at a sampling interval of 49.152\,$\mu$s. Bursts were identified for label construction using single-pulse pipelines based on PRESTO~\cite{Ransom01} and HEIMDALL~\cite{Petroff15}, with trial DMs of around the source's DM values. Events above an S/N threshold of 7 were retained and manually vetted. 

Conventional single-pulse searches are used only to assign supervised labels and to identify the raw-data intervals from which positive training segments are drawn. Search S/N, trial-DM curves and candidate rankings are not supplied as model inputs. Each constructed segment receives one of two labels. The positive class comprises astrophysical pulses from the two repeating FRBs and RRAT J1913+1330; the RRAT is included to broaden the training morphology of dispersed single pulses. The negative class comprises RFI and other non-astrophysical segments. All images related to the same physical pulse, neighbouring raw-data interval or observing unit are assigned to a single data partition before augmentation.

\subsubsection*{PSRFITS segmentation and dynamic-spectrum construction}\label{subsec:psrfits_pipeline}

The preprocessing stage constructs a dynamic-spectrum image from each selected interval of a PSRFITS observation~\cite{Hotan04}. The PSRFITS reader loads the channelized intensity array, which we denote by $I(c,s)$, where $c$ indexes frequency channel and $s$ indexes time sample. Each search-mode PSRFITS file corresponds to one beam of one observation and is processed independently; no combination across beams is performed. Total intensity is formed from the polarization channels stored in the \texttt{SUBINT} table according to \texttt{POL\_TYPE}: for Stokes data the Stokes-$I$ channel is taken; for dual-polarization data (for example \texttt{AABBCRCI}) the two hands of polarization are averaged as $\tfrac{1}{2}(A+B)$; and for single-polarization data the single recorded channel is used unchanged. Narrowband interference is removed after the block averaging described below: channels whose summed power or whose temporal standard deviation deviates from the corresponding robust estimate by more than $3\sigma$ are identified by iterative sigma clipping and replaced by the mean of the retained channels. Bad channels are therefore substituted rather than masked, so that no missing, flagged or undefined values enter the image.

Within each file segment, the intensity array is block-averaged over $N_f$ adjacent frequency channels and $N_t$ consecutive time samples,
$$
\bar{I}(i,j)=\frac{1}{N_fN_t} \sum_{p=0}^{N_f-1}\sum_{q=0}^{N_t-1} I(iN_f+p,jN_t+q),
$$
where the default values are $N_f=16$ channels and $N_t=16$ samples. For the FAST setup with 4096 channels over 1000--1500\,MHz and a raw sampling interval of 49.152\,$\mu$s, these defaults produce 256 frequency bins with a channel width of approximately 1.953\,MHz and a binned time resolution of approximately 0.786\,ms.

Each block-averaged segment is rendered as a two-dimensional dynamic-spectrum image with frequency and time along its two axes. The channel order is flipped where necessary so that the lowest frequency lies at the bottom of the image. The image is contrast-stretched by clipping the binned intensities of that segment to the interval $[\mu-3\sigma,\;\mu+3\sigma]$, where $\mu$ and $\sigma$ are the mean and standard deviation of that segment's own binned intensities, and mapping this interval linearly onto a single-channel greyscale ramp in which low values are white and high values are black; nearest-neighbour resampling is used, and the result is written as a PNG file. The image is then resized to $288\times288$ pixels before it is read by the model data loader. 
No de-dispersion or additional arrival-time alignment is applied before image construction, preserving the frequency-dependent pulse structure. Positive and negative segments are drawn from intervals identified by conventional searches and assigned labels through manual vetting. The model receives only the resulting image; search S/N, trial-DM curves and candidate rankings are not supplied as input features.

\subsubsection*{Parameterized quantum feature maps}\label{subsec:quantum_prelim}

The image produced from each raw-data segment is loaded using the same resizing and normalization procedure used during training. QViT returns a two-class probability vector, and the segment is flagged when its FRB-like probability exceeds a threshold fixed using the validation data.
QViT uses parameterized quantum circuits (PQCs) as trainable feature maps within the attention module. For a head-level input vector $\mathbf{x}\in\mathbb{R}^{d_h}$, an $n_q=d_h$ qubit register is initialized in $|0\rangle^{\otimes n_q}$, transformed by a data-encoding unitary $U_{\mathrm{enc}}(\mathbf{x})$ and a trainable unitary $U(\boldsymbol{\theta})$, and measured in the Pauli-$Z$ basis,
$$
|\psi(\mathbf{x};\boldsymbol{\theta})\rangle
=U(\boldsymbol{\theta})U_{\mathrm{enc}}(\mathbf{x})|0\rangle^{\otimes n_q},
\qquad
z_m=\langle\psi|Z_m|\psi\rangle.
$$
The measurement vector $\mathbf{z}=[z_1,\ldots,z_{n_q}]$ is therefore a dimension-preserving real-valued transformation from $\mathbb{R}^{d_h}$ to $\mathbb{R}^{d_h}$. The quantum circuit is not a stand-alone classifier; it replaces one projection layer inside an otherwise hybrid quantum--classical attention model.

\subsubsection*{ViT baseline}

The classical ViT follows the same tokenization scheme and overall encoder framework as QViT, while using classical linear Q/K/V projections. For both model families, the corresponding architectural parameters were varied over the same predefined ranges, and each configuration was trained using the same optimization protocol. All evaluated configurations are reported, while the configuration with the highest observed test-set recall within each model family is highlighted for descriptive comparison. Because these peak-recall configurations were identified independently, their embedding dimensions, projection dimensions, and parameter counts may differ.
It is deliberately constrained in a hardware-compatible manner: attention and embedding dimensions are matched to the quantum implementation, and parameter counts in the projection modules are matched where possible or reported explicitly where exact matching is not possible. The input dynamic spectrum is denoted by $\mathbf{Z}_0\in\mathbb{R}^{1\times H\times W}$, where $(H,W)$ is the image resolution.

\noindent\textbf{Patch embedding and tokenization.}
The implementation uses two convolutional projections. The first projection has kernel and stride $(h_1,w_1)$ and maps $\mathbf{Z}_0$ to a $d$-channel feature map $\mathbf{Z}_1\in\mathbb{R}^{d\times m_1\times n_1}$, where $m_1=H/h_1$ and $n_1=W/w_1$. A second convolution with kernel and stride $(k,k)$ merges each non-overlapping $k\times k$ neighbourhood and returns $\mathbf{Z}_2\in\mathbb{R}^{d\times m_2\times n_2}$, with $m_2=m_1/k$ and $n_2=n_1/k$. Flattening the spatial axes gives $N_0=m_2n_2$ image tokens $\mathbf{Z}_3\in\mathbb{R}^{N_0\times d}$. A learnable classification token $\mathbf{t}_{\mathrm{cls}}\in\mathbb{R}^{d}$ is prepended and a positional embedding $\mathbf{E}_{\mathrm{pos}}\in\mathbb{R}^{N\times d}$ is added,
$$
\mathbf{X}_0=[\mathbf{t}_{\mathrm{cls}};\mathbf{Z}_3]+\mathbf{E}_{\mathrm{pos}},
$$
where $N=N_0+1$. The resulting sequence is propagated through $L$ transformer encoder blocks.

\noindent\textbf{Multi-head self-attention.}
Let $\mathbf{X}_{\ell}\in\mathbb{R}^{N\times d}$ be the token sequence entering the $\ell$-th encoder block. The classical transformer block is
$$
\mathbf{Y}_{\ell}=\mathbf{X}_{\ell}+\mathrm{MSA}(\mathrm{Norm}(\mathbf{X}_{\ell})),\quad
\mathbf{X}_{\ell+1}=\mathbf{Y}_{\ell}+\mathrm{MLP}(\mathrm{Norm}(\mathbf{Y}_{\ell})),
$$
where $\mathrm{Norm}(\cdot)$ is the RMS normalization used in the implementation and $\mathrm{MLP}(\cdot)$ is a multi-layer perceptron with a GELU nonlinearity. Given $\mathbf{U}=\mathrm{Norm}(\mathbf{X}_{\ell})$, multi-head self-attention forms query, key and value matrices by three learned linear projections,
$$
\mathbf{Q}=\mathbf{U}\mathbf{W}_Q,\quad
\mathbf{K}=\mathbf{U}\mathbf{W}_K,\quad
\mathbf{V}=\mathbf{U}\mathbf{W}_V,
$$
with $\mathbf{W}_Q,\mathbf{W}_K,\mathbf{W}_V\in\mathbb{R}^{d\times d_{\mathrm{att}}}$, $d_{\mathrm{att}}=n_h d_h$ and $n_h$ the number of attention heads. These are split into $n_h$ heads, $\mathbf{Q}^{(i)},\mathbf{K}^{(i)},\mathbf{V}^{(i)}\in\mathbb{R}^{N\times d_h}$, each computing scaled dot-product attention
$$
\mathbf{A}^{(i)}=\mathrm{Softmax}\left(\frac{\mathbf{Q}^{(i)}(\mathbf{K}^{(i)})^\top}{\sqrt{d_h}}\right)\mathbf{V}^{(i)},
$$
and the concatenated head outputs are projected back to the embedding dimension via
$$
\mathrm{MSA}(\mathbf{U})=\mathbf{A}\mathbf{W}_O+\mathbf{b}_O,
$$
with $\mathbf{A}=[\mathbf{A}^{(1)},\ldots,\mathbf{A}^{(n_h)}]\in\mathbb{R}^{N\times d_{\mathrm{att}}}$, $\mathbf{W}_O\in\mathbb{R}^{d_{\mathrm{att}}\times d}$, $\mathbf{b}_O\in\mathbb{R}^{d}$. 
The same encoder structure and output projection $\mathbf{W}_O$ are used by both models, so the only difference between the two architectures lies in the projection module.

\noindent\textbf{Classification head.}
The full model is $\mathbf{p}=\mathrm{Softmax}(\mathrm{Classifier}(\mathbf{o}_{\mathrm{cls}}))$, where $\mathbf{o}_{\mathrm{cls}}\in\mathbb{R}^{d}$ is the output of the classification token after the final block, and
$$
\mathrm{Classifier}(\mathbf{o}_{\mathrm{cls}})=\mathbf{o}_{\mathrm{cls}}\mathbf{W}_{\mathrm{cls}}+\mathbf{b}_{\mathrm{cls}},
$$
with $\mathbf{W}_{\mathrm{cls}}\in\mathbb{R}^{d\times C_{\mathrm{cls}}}$, $\mathbf{b}_{\mathrm{cls}}\in\mathbb{R}^{C_{\mathrm{cls}}}$ and $\mathbf{p}=[p_c]_{c\in\mathcal{C}}$. 
QViT and ViT use the same functional form for the classification head, with the input dimension matched to the embedding dimension $d$ of each respective configuration and the same output dimension $C_{\mathrm{cls}}$.

\subsubsection*{Quantum multi-head self-attention}
Let $\mathbf{U}\in\mathbb{R}^{N\times d}$ denote the normalized token matrix entering $\mathrm{QMSA}(\cdot)$. QViT keeps the usual multi-head organization, but replaces the learned linear maps that generate queries, keys and values with parameterized quantum circuits. The embedding dimension is split into $n_h$ heads, $\mathbf{U}=[\mathbf{U}^{(1)},\ldots,\mathbf{U}^{(n_h)}]$ with $\mathbf{U}^{(i)}\in\mathbb{R}^{N\times d_h}$ and $d=n_h d_h$. For the $i$-th head, three independent quantum projections generate the query, key and value representations,
$$
\mathbf{Q}^{(i)}=\mathrm{QProj}_{Q}(\mathbf{U}^{(i)}),\quad
\mathbf{K}^{(i)}=\mathrm{QProj}_{K}(\mathbf{U}^{(i)}),\quad
\mathbf{V}^{(i)}=\mathrm{QProj}_{V}(\mathbf{U}^{(i)}),
$$
each applied row by row. With angle encoding, a head-level token vector of length $d_h$ is encoded on $n_q=d_h$ qubits and Pauli-$Z$ measurements return one real number per qubit, so the quantum projection maps $\mathbb{R}^{d_h}$ to $\mathbb{R}^{d_h}$. Because the circuit is dimension-preserving, the head outputs already have dimension $d$, and QViT uses the same output projection $\mathbf{W}_O$ as the ViT baseline, so the two architectures differ only in the projection module.

\noindent\textbf{Enhanced parameterized quantum circuit.}
The enhanced PQC adds a trainable enhancement block before data encoding, followed by repeated variational layers. Each circuit starts from $|0\rangle^{\otimes n_q}$. The enhancement unitary is
$$
U_{\mathrm{enh}}=U_{\mathrm{ent}}^{(0)}U_{\mathrm{rot}}^{(0)},\quad
U_{\mathrm{rot}}^{(0)}=\bigotimes_{m=1}^{n_q}R_Y(\alpha_m)R_Z(\beta_m),\quad
U_{\mathrm{ent}}^{(0)}=\prod_{m=1}^{n_q-1}CZ_{m,m+1},
$$
with trainable $\alpha_m$, $\beta_m$, and $CZ_{m,m+1}$ a controlled-$Z$ gate between adjacent qubits. The input encoding unitary is
$$
U_{\mathrm{enc}}(\mathbf{x}_j)=\bigotimes_{m=1}^{n_q}R_X(x_{j,m}),
$$
and the variational layer is $U_{\mathrm{layer}}=U_{\mathrm{ent}}U_{\mathrm{rot}}$ with
$$
U_{\mathrm{rot}}=\bigotimes_{m=1}^{n_q}R_Y(\phi_m)R_Z(\psi_m),\quad
U_{\mathrm{ent}}=\prod_{m=1}^{n_q-1}CZ_{m,m+1}.
$$
For circuit depth $L_q$, the repeated variational unitary is $U_{\mathrm{PQC}}=\prod_{\ell=1}^{L_q}U_{\mathrm{layer}}^{(\ell)}$, and the full circuit is
$$
\mathrm{PQC}_{\star}(\mathbf{x}_j;\boldsymbol{\theta}_{\star})=U_{\mathrm{PQC}}U_{\mathrm{enc}}(\mathbf{x}_j)U_{\mathrm{enh}}|0\rangle^{\otimes n_q},
$$
where operators act from right to left. After preparing $|\psi\rangle$, each qubit is measured in the Pauli-$Z$ basis, returning $\mathbf{z}=[z_1,\dots,z_{n_q}]\in\mathbb{R}^{n_q}$ with $z_m=\langle\psi|Z_m|\psi\rangle$. Parameter-budget sweeps modify only the trainable gates inside $\mathrm{PQC}_{\star}(\cdot)$; tokenization, attention-head structure and the classifier are unchanged.

\subsubsection*{Evaluation metrics and hardware consistency}
\noindent\textbf{Accuracy and recall:}
The model makes a signal/background decision per dynamic-spectrum segment. With $TP$, $TN$, $FP$ and $FN$ the true/false positives and negatives,
$$
\mathrm{Acc}=\frac{TP+TN}{TP+TN+FP+FN},\quad
\mathrm{Recall}=\frac{TP}{TP+FN}.
$$
These quantities evaluate classification of labelled segments. They should not be interpreted as event-level search completeness or false-positive rate per observing hour, which require blind scanning of continuous observations.

\noindent\textbf{Hellinger fidelity:}
To compare ideal noiseless simulation with quantum hardware output distributions we used the Hellinger fidelity,
$$
F_{\mathrm{H}}(p,q)=\left(\sum_i\sqrt{p_iq_i}\right)^2,
$$
where $p_i$ and $q_i$ are the probabilities of outcome $i$ from simulation and hardware. $F_{\mathrm{H}}\in[0,1]$, with values closer to 1 indicating closer agreement.

\noindent\textbf{Cosine similarity of measured quantum features:}
We also computed the cosine similarity between simulated and real-device measurement vectors,
$$
S_{\cos}(\mathbf{s},\mathbf{r})=\frac{\mathbf{s}\cdot\mathbf{r}}{\|\mathbf{s}\|_2\|\mathbf{r}\|_2},
$$
where $\mathbf{s}$ and $\mathbf{r}$ are the measured Pauli-$Z$ expectation-value vectors from simulation and real-device execution; a value close to 1 indicates that the device returns a feature vector aligned with the simulator output.

\subsubsection*{Statistical analysis}

Model performance is summarized over independent random initializations using the mean and standard deviation. QViT and ViT are evaluated on identical data partitions, with the decision threshold fixed using validation data. Because multiple segments may originate from the same pulse or observation, segment-level predictions are not assumed to be mutually independent. No confidence intervals or hypothesis tests are reported here; the reported standard deviations describe run-to-run variability across the 10 seeds.

The QViT patch-embedding module uses a convolutional stem with total stride $(48,48)$ for the $288\times288$ input, producing a $6\times6$ patch grid, or 36 image tokens; after adding the CLS token the sequence length is $N=37$. The number of attention heads was fixed at $n_h=4$. The training script uses a one-block encoder, learnable positional embeddings, RMS normalization, GELU activation in the MLP, an MLP expansion ratio 6.0 and linear dropout probability 0.1. Each configuration was repeated over 10 independent random seeds. Training used a batch size of 32 and 120 epochs, with AdamW (learning rate $4\times10^{-4}$, weight decay $10^{-2}$), an 8-epoch warm-up followed by cosine decay, Focal loss with $\gamma=2.0$ and label smoothing 0.005, and best-checkpoint selection by validation-set FRB-class $F_2$.

\begin{figure*}
\centering
\includegraphics[width=\linewidth]{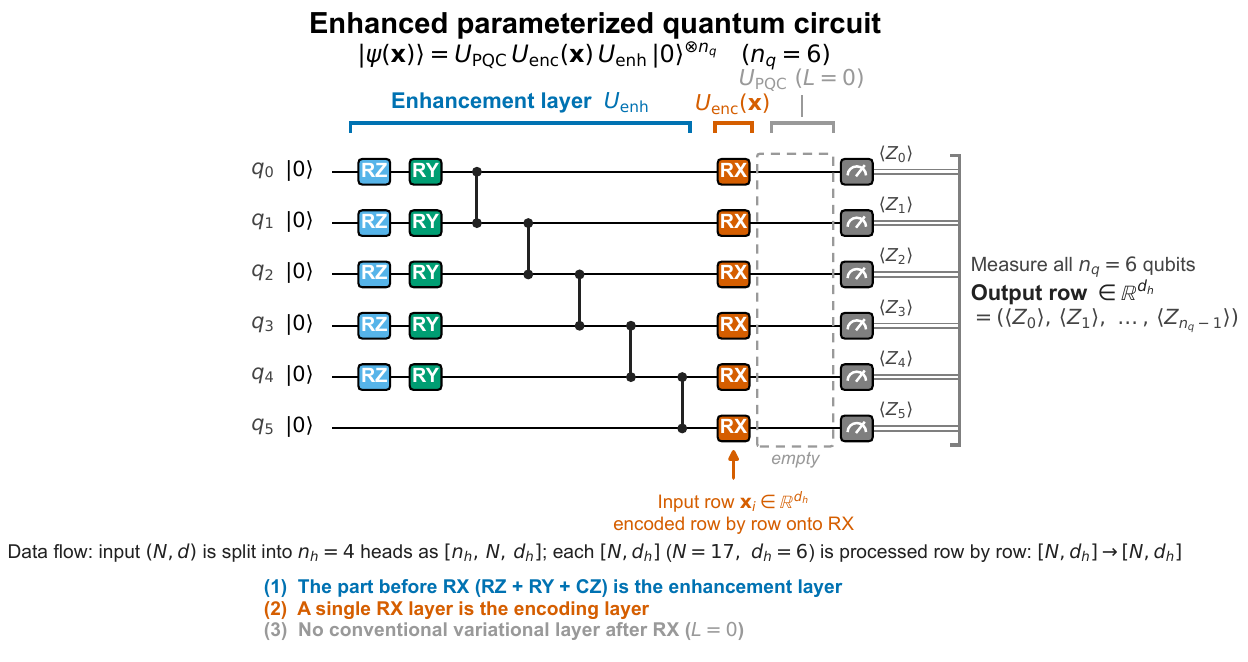}
\caption{Enhanced parameterized quantum circuit used for the Q/K/V projections and for real-device inference.}
\label{fig_real_qc_used}
\end{figure*}

For real-device validation, we used the separate hardware-compatible configuration shown in Figure~\ref{fig_real_qc_used}: $n_q=d_h=6$, $n_h=4$, $N=17$ and $L_q=0$. The three Q/K/V projections across the four heads therefore require $3\times4\times17=204$ circuit evaluations per image. This configuration is distinct from the four-qubit, one-variational-layer, $N=37$ model used for the comparison in Table~\ref{tab:main_performance}.

\clearpage
\bibliography{biblio.bib}

@ARTICLE{Kordzanganeh21_QML_RA,
       author = {{Kordzanganeh}, Mohammad and {Utting}, Aydin and {Scaife}, Anna},
        title = "{Quantum Machine Learning for Radio Astronomy}",
      journal = {arXiv e-prints},
         year = 2021,
        month = dec,
          eid = {arXiv:2112.02655},
        pages = {arXiv:2112.02655},
          doi = {10.48550/arXiv.2112.02655},
archivePrefix = {arXiv},
       eprint = {2112.02655},
 primaryClass = {quant-ph},
       adsurl = {https://ui.adsabs.harvard.edu/abs/2021arXiv211202655K}
}

@ARTICLE{Zhang25_HQViT,
       author = {{Zhang}, Hui and {Zhao}, Qinglin and {Zhou}, Mengchu and {Feng}, Li},
        title = "{HQViT: Hybrid Quantum Vision Transformer for Image Classification}",
      journal = {arXiv e-prints},
         year = 2025,
        month = apr,
          eid = {arXiv:2504.02730},
        pages = {arXiv:2504.02730},
          doi = {10.48550/arXiv.2504.02730},
archivePrefix = {arXiv},
       eprint = {2504.02730},
 primaryClass = {cs.CV},
       adsurl = {https://ui.adsabs.harvard.edu/abs/2025arXiv250402730Z}
}

@ARTICLE{Cherrat24_QViT,
       author = {{Cherrat}, El Amine and {Kerenidis}, Iordanis and {Mathur}, Natansh and {Landman}, Jonas and {Strahm}, Martin and {Li}, Yun Yvonna},
        title = "{Quantum Vision Transformers}",
      journal = {Quantum},
         year = 2024,
        month = feb,
       volume = {8},
        pages = {1265},
          doi = {10.22331/q-2024-02-22-1265},
archivePrefix = {arXiv},
       eprint = {2209.08167},
 primaryClass = {quant-ph},
       adsurl = {https://ui.adsabs.harvard.edu/abs/2024Quant...8.1265C}
}

@ARTICLE{Dewdney09_SKA,
       author = {{Dewdney}, P.~E. and {Hall}, P.~J. and {Schilizzi}, R.~T. and {Lazio}, T.~J.~L.~W.},
        title = "{The Square Kilometre Array}",
      journal = {IEEE Proceedings},
         year = 2009,
        month = aug,
       volume = {97},
       number = {8},
        pages = {1482-1496},
          doi = {10.1109/JPROC.2009.2021005},
       adsurl = {https://ui.adsabs.harvard.edu/abs/2009IEEEP..97.1482D}
}

@ARTICLE{Connor26_CASM,
       author = {{Connor}, Liam and {Ravi}, Vikram and {Sanghavi}, Pranav and {Balakrishan}, Vishnu and {Chung}, Luke and {Daghlian}, Saren and {Dunn}, Liam and {Griffin}, Anthony and {Harnach}, Charlie and {Hodges}, Mark and {Jameson}, Andrew and {Gutierrez}, Michael and {Leung}, Calvin and {Lin}, Mei and {Mehla}, Advait and {Modilim}, Obinna and {Patel}, Nimesh and {Smith}, Kendrick and {Zeng}, Lingzhen},
        title = "{The 256-antenna Coherent All-Sky Monitor}",
      journal = {arXiv e-prints},
         year = 2026,
        month = apr,
          eid = {arXiv:2604.13903},
        pages = {arXiv:2604.13903},
          doi = {10.48550/arXiv.2604.13903},
archivePrefix = {arXiv},
       eprint = {2604.13903},
 primaryClass = {astro-ph.IM},
       adsurl = {https://ui.adsabs.harvard.edu/abs/2026arXiv260413903C}
}

@ARTICLE{Jiang24_FASTCA,
       author = {{Jiang}, Peng and {Chen}, Rurong and {Gan}, Hengqian and {Sun}, Jinghai and {Zhu}, Boqin and {Li}, Hui and {Zhu}, Weiwei and {Wu}, Jingwen and {Chen}, Xuelei and {Zhang}, Haiyan and {An}, Tao},
        title = "{The FAST Core Array}",
      journal = {Astronomical Techniques and Instruments},
         year = 2024,
        month = jan,
       volume = {1},
       number = {2},
        pages = {84-94},
          doi = {10.61977/ati2024012},
archivePrefix = {arXiv},
       eprint = {2408.12826},
 primaryClass = {astro-ph.IM},
       adsurl = {https://ui.adsabs.harvard.edu/abs/2024AstTI...1...84J}
}

@ARTICLE{Zhang23_0912A,
       author = {{Zhang}, Yong-Kun and {Li}, Di and {Zhang}, Bing and {Cao}, Shuo and {Feng}, Yi and {Wang}, Wei-Yang and {Qu}, Yuanhong and {Niu}, Jia-Rui and {Zhu}, Wei-Wei and {Han}, Jin-Lin and {Jiang}, Peng and {Lee}, Ke-Jia and {Li}, Dong-Zi and {Luo}, Rui and {Niu}, Chen-Hui and {Tsai}, Chao-Wei and {Wang}, Pei and {Wang}, Fa-Yin and {Wu}, Zi-Wei and {Xu}, Heng and {Yang}, Yuan-Pei and {Zhang}, Jun-Shuo and {Zhou}, De-Jiang and {Zhu}, Yu-Hao},
        title = "{FAST Observations of FRB 20220912A: Burst Properties and Polarization Characteristics}",
      journal = {\apj},
         year = 2023,
        month = oct,
       volume = {955},
       number = {2},
          eid = {142},
        pages = {142},
          doi = {10.3847/1538-4357/aced0b},
archivePrefix = {arXiv},
       eprint = {2304.14665},
 primaryClass = {astro-ph.HE},
       adsurl = {https://ui.adsabs.harvard.edu/abs/2023ApJ...955..142Z}
}

@ARTICLE{Yang26,
       author = {{Yang}, X. and {Zhang}, S.~B. and {Li}, Y. and {Xiao}, D. and {Zhang}, W.~L. and {Wei}, J.-J. and {Geng}, J.-J. and {Wang}, J.-S. and {Yang}, Y.~P. and {Wang}, F.~Y. and {Wu}, X.~F. and {Dai}, Z.~G.},
        title = "{An invariant energy release hierarchy in a repeating fast radio burst}",
      journal = {arXiv e-prints},
         year = 2026,
        month = aug,
          eid = {arXiv:2608.18455},
        pages = {arXiv:2608.18455},
          doi = {10.48550/arXiv.2608.18455},
archivePrefix = {arXiv},
       eprint = {2608.18455},
 primaryClass = {astro-ph.HE},
       adsurl = {https://ui.adsabs.harvard.edu/abs/2026arXiv260818455Y}
}

@ARTICLE{McLaughlin06,
       author = {{McLaughlin}, M.~A. and {Lyne}, A.~G. and {Lorimer}, D.~R. and {Kramer}, M. and {Faulkner}, A.~J. and {Manchester}, R.~N. and {Cordes}, J.~M. and {Camilo}, F. and {Possenti}, A. and {Stairs}, I.~H. and {Hobbs}, G. and {D'Amico}, N. and {Burgay}, M. and {O'Brien}, J.~T.},
        title = "{Transient radio bursts from rotating neutron stars}",
      journal = {\nat},
         year = 2006,
        month = feb,
       volume = {439},
       number = {7078},
        pages = {817-820},
          doi = {10.1038/nature04440},
archivePrefix = {arXiv},
       eprint = {astro-ph/0511587},
 primaryClass = {astro-ph},
       adsurl = {https://ui.adsabs.harvard.edu/abs/2006Natur.439..817M}
}

@ARTICLE{Chen26,
       author = {{Chen}, Yunchuan and {Ni}, Shulei and {Li}, Chan and {Fang}, Jianhua and {Zhou}, Dengke and {Chen}, Huaxi and {Feng}, Yi and {Wang}, Pei and {Jin}, Chenwu and {Wang}, Han and {Huang}, Bijuan and {Guo}, Xuerong and {Quan}, Donghui and {Li}, Di},
        title = "{SwinYNet: A Transformer-based Multitask Model for Accurate and Efficient Fast Radio Burst Searches}",
      journal = {\apjs},
         year = 2026,
        month = apr,
       volume = {283},
       number = {2},
          eid = {45},
        pages = {45},
          doi = {10.3847/1538-4365/ae40f7},
archivePrefix = {arXiv},
       eprint = {2603.05958},
 primaryClass = {astro-ph.GA},
       adsurl = {https://ui.adsabs.harvard.edu/abs/2026ApJS..283...45C}
}

@ARTICLE{Petroff19AARv,
       author = {{Petroff}, E. and {Hessels}, J.~W.~T. and {Lorimer}, D.~R.},
        title = "{Fast radio bursts}",
      journal = {\aapr},
         year = 2019,
        month = dec,
       volume = {27},
       number = {1},
          eid = {4},
        pages = {4},
          doi = {10.1007/s00159-019-0116-6},
archivePrefix = {arXiv},
       eprint = {1904.07947},
 primaryClass = {astro-ph.HE},
       adsurl = {https://ui.adsabs.harvard.edu/abs/2019A&ARv..27....4P}
}

@ARTICLE{Zhang20_PTD1,
       author = {{Zhang}, S.-B. and {Hobbs}, G. and {Russell}, C.~J. and {Toomey}, L. and {Dai}, S. and {Dempsey}, J. and {Manchester}, R.~N. and {Johnston}, S. and {Staveley-Smith}, L. and {Wu}, X.-F. and {Li}, D. and {Yang}, Y.-Y. and {Wang}, S.-Q. and {Qiu}, H. and {Luo}, R. and {Wang}, C. and {Zhang}, C. and {Zhang}, L. and {Mandow}, R.},
        title = "{Parkes Transient Events. I. Database of Single Pulses, Initial Results, and Missing Fast Radio Bursts}",
      journal = {\apjs},
         year = 2020,
        month = jul,
       volume = {249},
       number = {1},
          eid = {14},
        pages = {14},
          doi = {10.3847/1538-4365/ab95a4},
archivePrefix = {arXiv},
       eprint = {2004.04601},
 primaryClass = {astro-ph.HE},
       adsurl = {https://ui.adsabs.harvard.edu/abs/2020ApJS..249...14Z}
}

@ARTICLE{Yang21,
       author = {{Yang}, X. and {Zhang}, S.-B. and {Wang}, J.-S. and {Hobbs}, G. and {Sun}, T.-R. and {Manchester}, R.~N. and {Geng}, J.-J. and {Russell}, C.~J. and {Luo}, R. and {Tang}, Z.-F. and {Wang}, C. and {Wei}, J.-J. and {Staveley-Smith}, L. and {Dai}, S. and {Li}, Y. and {Yang}, Y.-Y. and {Wu}, X.-F.},
        title = "{81 New candidate fast radio bursts in Parkes archive}",
      journal = {\mnras},
         year = 2021,
        month = nov,
       volume = {507},
       number = {3},
        pages = {3238-3245},
          doi = {10.1093/mnras/stab2275},
archivePrefix = {arXiv},
       eprint = {2108.00609},
 primaryClass = {astro-ph.HE},
       adsurl = {https://ui.adsabs.harvard.edu/abs/2021MNRAS.507.3238Y}
}

@ARTICLE{Fetch,
       author = {{Agarwal}, Devansh and {Aggarwal}, Kshitij and {Burke-Spolaor}, Sarah and {Lorimer}, Duncan R. and {Garver-Daniels}, Nathaniel},
        title = "{FETCH: A deep-learning based classifier for fast transient classification}",
      journal = {\mnras},
         year = 2020,
        month = sep,
       volume = {497},
       number = {2},
        pages = {1661-1674},
          doi = {10.1093/mnras/staa1856},
archivePrefix = {arXiv},
       eprint = {1902.06343},
 primaryClass = {astro-ph.IM},
       adsurl = {https://ui.adsabs.harvard.edu/abs/2020MNRAS.497.1661A}
}

@ARTICLE{Cordes03,
       author = {{Cordes}, J.~M. and {McLaughlin}, M.~A.},
        title = "{Searches for Fast Radio Transients}",
      journal = {\apj},
         year = 2003,
        month = oct,
       volume = {596},
       number = {2},
        pages = {1142-1154},
          doi = {10.1086/378231},
archivePrefix = {arXiv},
       eprint = {astro-ph/0304364},
 primaryClass = {astro-ph},
       adsurl = {https://ui.adsabs.harvard.edu/abs/2003ApJ...596.1142C}
}

@ARTICLE{Li26,
       author = {{Li}, Y. and {Zhang}, S.~B. and {Yang}, Y.~P. and {Tsai}, C.~W. and {Yang}, X. and {Law}, C.~J. and {Anna-Thomas}, R. and {Chen}, X.~L. and {Lee}, K.~J. and {Tang}, Z.~F. and {Xiao}, D. and {Xu}, H. and {Yang}, X.~L. and {Chen}, G. and {Feng}, Y. and {Li}, D.~Z. and {Mckinven}, R. and {Niu}, J.~R. and {Shin}, K. and {Wang}, B.~J. and {Zhang}, C.~F. and {Zhang}, Y.~K. and {Zhou}, D.~J. and {Zhu}, Y.~H. and {Dai}, Z.~G. and {Chang}, C.~M. and {Geng}, J.~J. and {Han}, J.~L. and {Hu}, L. and {Li}, D. and {Luo}, R. and {Niu}, C.~H. and {Shi}, D.~D. and {Sun}, T.~R. and {Wu}, X.~F. and {Zhu}, W.~W. and {Jiang}, P. and {Zhang}, B.},
        title = "{A sudden change and recovery in the magnetic environment around a repeating fast radio burst}",
      journal = {Science},
         year = 2026,
        month = jan,
       volume = {391},
       number = {6782},
        pages = {280-284},
          doi = {10.1126/science.adq3225},
archivePrefix = {arXiv},
       eprint = {2503.04727},
 primaryClass = {astro-ph.HE},
       adsurl = {https://ui.adsabs.harvard.edu/abs/2026Sci...391..280L}
}

@ARTICLE{Niu21,
       author = {{Niu}, Chen-Hui and {Li}, Di and {Luo}, Rui and {Wang}, Wei-Yang and {Yao}, Jumei and {Zhang}, Bing and {Zhu}, Wei-Wei and {Wang}, Pei and {Ye}, Haoyang and {Zhang}, Yong-Kun and {Niu}, Jia-rui and {Tang}, Ning-yu and {Duan}, Ran and {Krco}, Marko and {Dai}, Shi and {Feng}, Yi and {Miao}, Chenchen and {Pan}, Zhichen and {Qian}, Lei and {Xue}, Mengyao and {Yuan}, Mao and {Yue}, Youling and {Zhang}, Lei and {Zhang}, Xinxin},
        title = "{CRAFTS for Fast Radio Bursts: Extending the Dispersion-Fluence Relation with New FRBs Detected by FAST}",
      journal = {\apjl},
         year = 2021,
        month = mar,
       volume = {909},
       number = {1},
          eid = {L8},
        pages = {L8},
          doi = {10.3847/2041-8213/abe7f0},
archivePrefix = {arXiv},
       eprint = {2102.10546},
 primaryClass = {astro-ph.HE},
       adsurl = {https://ui.adsabs.harvard.edu/abs/2021ApJ...909L...8N}
}

@ARTICLE{Lorimer07,
       author = {{Lorimer}, D.~R. and {Bailes}, M. and {McLaughlin}, M.~A. and {Narkevic}, D.~J. and {Crawford}, F.},
        title = "{A Bright Millisecond Radio Burst of Extragalactic Origin}",
      journal = {Science},
         year = 2007,
        month = nov,
       volume = {318},
       number = {5851},
        pages = {777},
          doi = {10.1126/science.1147532},
archivePrefix = {arXiv},
       eprint = {0709.4301},
 primaryClass = {astro-ph},
       adsurl = {https://ui.adsabs.harvard.edu/abs/2007Sci...318..777L}
}

@ARTICLE{Thornton13,
       author = {{Thornton}, D. and {Stappers}, B. and {Bailes}, M. and {Barsdell}, B. and {Bates}, S. and {Bhat}, N.~D.~R. and {Burgay}, M. and {Burke-Spolaor}, S. and {Champion}, D.~J. and {Coster}, P. and {D'Amico}, N. and {Jameson}, A. and {Johnston}, S. and {Keith}, M. and {Kramer}, M. and {Levin}, L. and {Milia}, S. and {Ng}, C. and {Possenti}, A. and {van Straten}, W.},
        title = "{A Population of Fast Radio Bursts at Cosmological Distances}",
      journal = {Science},
         year = 2013,
        month = jul,
       volume = {341},
       number = {6141},
        pages = {53-56},
          doi = {10.1126/science.1236789},
archivePrefix = {arXiv},
       eprint = {1307.1628},
 primaryClass = {astro-ph.HE},
       adsurl = {https://ui.adsabs.harvard.edu/abs/2013Sci...341...53T}
}

@ARTICLE{Xu22,
       author = {{Xu}, H. and {Niu}, J.~R. and {Chen}, P. and {Lee}, K.~J. and {Zhu}, W.~W. and {Dong}, S. and {Zhang}, B. and {Jiang}, J.~C. and {Wang}, B.~J. and {Xu}, J.~W. and {Zhang}, C.~F. and {Fu}, H. and {Filippenko}, A.~V. and {Peng}, E.~W. and {Zhou}, D.~J. and {Zhang}, Y.~K. and {Wang}, P. and {Feng}, Y. and {Li}, Y. and {Brink}, T.~G. and {Li}, D.~Z. and {Lu}, W. and {Yang}, Y.~P. and {Caballero}, R.~N. and {Cai}, C. and {Chen}, M.~Z. and {Dai}, Z.~G. and {Djorgovski}, S.~G. and {Esamdin}, A. and {Gan}, H.~Q. and {Guhathakurta}, P. and {Han}, J.~L. and {Hao}, L.~F. and {Huang}, Y.~X. and {Jiang}, P. and {Li}, C.~K. and {Li}, D. and {Li}, H. and {Li}, X.~Q. and {Li}, Z.~X. and {Liu}, Z.~Y. and {Luo}, R. and {Men}, Y.~P. and {Niu}, C.~H. and {Peng}, W.~X. and {Qian}, L. and {Song}, L.~M. and {Stern}, D. and {Stockton}, A. and {Sun}, J.~H. and {Wang}, F.~Y. and {Wang}, M. and {Wang}, N. and {Wang}, W.~Y. and {Wu}, X.~F. and {Xiao}, S. and {Xiong}, S.~L. and {Xu}, Y.~H. and {Xu}, R.~X. and {Yang}, J. and {Yang}, X. and {Yao}, R. and {Yi}, Q.~B. and {Yue}, Y.~L. and {Yu}, D.~J. and {Yu}, W.~F. and {Yuan}, J.~P. and {Zhang}, B.~B. and {Zhang}, S.~B. and {Zhang}, S.~N. and {Zhao}, Y. and {Zheng}, W.~K. and {Zhu}, Y. and {Zou}, J.~H.},
        title = "{A fast radio burst source at a complex magnetized site in a barred galaxy}",
      journal = {\nat},
         year = 2022,
        month = sep,
       volume = {609},
       number = {7928},
        pages = {685-688},
          doi = {10.1038/s41586-022-05071-8},
archivePrefix = {arXiv},
       eprint = {2111.11764},
 primaryClass = {astro-ph.HE},
       adsurl = {https://ui.adsabs.harvard.edu/abs/2022Natur.609..685X}
}

@ARTICLE{Li21,
       author = {{Li}, D. and {Wang}, P. and {Zhu}, W.~W. and {Zhang}, B. and {Zhang}, X.~X. and {Duan}, R. and {Zhang}, Y.~K. and {Feng}, Y. and {Tang}, N.~Y. and {Chatterjee}, S. and {Cordes}, J.~M. and {Cruces}, M. and {Dai}, S. and {Gajjar}, V. and {Hobbs}, G. and {Jin}, C. and {Kramer}, M. and {Lorimer}, D.~R. and {Miao}, C.~C. and {Niu}, C.~H. and {Niu}, J.~R. and {Pan}, Z.~C. and {Qian}, L. and {Spitler}, L. and {Werthimer}, D. and {Zhang}, G.~Q. and {Wang}, F.~Y. and {Xie}, X.~Y. and {Yue}, Y.~L. and {Zhang}, L. and {Zhi}, Q.~J. and {Zhu}, Y.},
        title = "{A bimodal burst energy distribution of a repeating fast radio burst source}",
      journal = {\nat},
         year = 2021,
        month = oct,
       volume = {598},
       number = {7880},
        pages = {267-271},
          doi = {10.1038/s41586-021-03878-5},
archivePrefix = {arXiv},
       eprint = {2107.08205},
 primaryClass = {astro-ph.HE},
       adsurl = {https://ui.adsabs.harvard.edu/abs/2021Natur.598..267L}
}

@ARTICLE{Hotan04,
       author = {{Hotan}, A.~W. and {van Straten}, W. and {Manchester}, R.~N.},
        title = "{PSRCHIVE and PSRFITS: An Open Approach to Radio Pulsar Data Storage and Analysis}",
      journal = {\pasa},
         year = 2004,
        month = jan,
       volume = {21},
       number = {3},
        pages = {302-309},
          doi = {10.1071/AS04022},
archivePrefix = {arXiv},
       eprint = {astro-ph/0404549},
 primaryClass = {astro-ph},
       adsurl = {https://ui.adsabs.harvard.edu/abs/2004PASA...21..302H}
}

@PHDTHESIS{Ransom01,
       author = {{Ransom}, Scott Mitchell},
        title = "{New search techniques for binary pulsars}",
       school = {Harvard University, Massachusetts},
         year = 2001,
        month = jan,
       adsurl = {https://ui.adsabs.harvard.edu/abs/2001PhDT.......123R}
}

@ARTICLE{Petroff15,
       author = {{Petroff}, E. and {Bailes}, M. and {Barr}, E.~D. and {Barsdell}, B.~R. and {Bhat}, N.~D.~R. and {Bian}, F. and {Burke-Spolaor}, S. and {Caleb}, M. and {Champion}, D. and {Chandra}, P. and {Da Costa}, G. and {Delvaux}, C. and {Flynn}, C. and {Gehrels}, N. and {Greiner}, J. and {Jameson}, A. and {Johnston}, S. and {Kasliwal}, M.~M. and {Keane}, E.~F. and {Keller}, S. and {Kocz}, J. and {Kramer}, M. and {Leloudas}, G. and {Malesani}, D. and {Mulchaey}, J.~S. and {Ng}, C. and {Ofek}, E.~O. and {Perley}, D.~A. and {Possenti}, A. and {Schmidt}, B.~P. and {Shen}, Yue and {Stappers}, B. and {Tisserand}, P. and {van Straten}, W. and {Wolf}, C.},
        title = "{A real-time fast radio burst: polarization detection and multiwavelength follow-up}",
      journal = {\mnras},
         year = 2015,
        month = feb,
       volume = {447},
       number = {1},
        pages = {246-255},
          doi = {10.1093/mnras/stu2419},
archivePrefix = {arXiv},
       eprint = {1412.0342},
 primaryClass = {astro-ph.HE},
       adsurl = {https://ui.adsabs.harvard.edu/abs/2015MNRAS.447..246P}
}

@article{Spitler16,
    author = "Spitler, L.G. and others",
    title = "{A Repeating Fast Radio Burst}",
    eprint = "1603.00581",
    archivePrefix = "arXiv",
    primaryClass = "astro-ph.HE",
    doi = "10.1038/nature17168",
    journal = "Nature",
    volume = "531",
    pages = "202",
    year = "2016"
}

@ARTICLE{Jiang20,
       author = {{Jiang}, Peng and {Tang}, Ning-Yu and {Hou}, Li-Gang and {Liu}, Meng-Ting and {Kr{\v{c}}o}, Marko and {Qian}, Lei and {Sun}, Jing-Hai and {Ching}, Tao-Chung and {Liu}, Bin and {Duan}, Yan and {Yue}, You-Ling and {Gan}, Heng-Qian and {Yao}, Rui and {Li}, Hui and {Pan}, Gao-Feng and {Yu}, Dong-Jun and {Liu}, Hong-Fei and {Li}, Di and {Peng}, Bo and {Yan}, Jun and {FAST Collaboration}},
        title = "{The fundamental performance of FAST with 19-beam receiver at L band}",
      journal = {Research in Astronomy and Astrophysics},
         year = 2020,
        month = may,
       volume = {20},
       number = {5},
          eid = {064},
        pages = {064},
          doi = {10.1088/1674-4527/20/5/64},
archivePrefix = {arXiv},
       eprint = {2002.01786},
 primaryClass = {astro-ph.IM},
       adsurl = {https://ui.adsabs.harvard.edu/abs/2020RAA....20...64J}
}

@ARTICLE{Zhang23_1913,
       author = {{Zhang}, S.~B. and {Geng}, J.~J. and {Wang}, J.~S. and {Yang}, X. and {Kaczmarek}, J. and {Tang}, Z.~F. and {Johnston}, S. and {Hobbs}, G. and {Manchester}, R. and {Wu}, X.~F. and {Jiang}, P. and {Huang}, Y.~F. and {Zou}, Y.~C. and {Dai}, Z.~G. and {Zhang}, B. and {Li}, D. and {Yang}, Y.~P. and {Dai}, S. and {Chang}, C.~M. and {Pan}, Z.~C. and {Lu}, J.~G. and {Wei}, J.~J. and {Li}, Y. and {Wu}, Q.~W. and {Qian}, L. and {Wang}, P. and {Wang}, S.~Q. and {Feng}, Y. and {Staveley-Smith}, L.},
        title = "{RRAT J1913+1330: an extremely variable and puzzling pulsar}",
      journal = {arXiv e-prints},
         year = 2023,
        month = jun,
          eid = {arXiv:2306.02855},
        pages = {arXiv:2306.02855},
          doi = {10.48550/arXiv.2306.02855},
archivePrefix = {arXiv},
       eprint = {2306.02855},
 primaryClass = {astro-ph.HE},
       adsurl = {https://ui.adsabs.harvard.edu/abs/2023arXiv230602855Z}
}

@ARTICLE{dosovitskiy2020,
      author = {{Dosovitskiy}, A. and {Beyer}, L. and {Kolesnikov}, A. and {Weissenborn}, D. and {Zhai}, X. and {Unterthiner}, T. and {Dehghani}, M. and {Minderer}, M. and {Heigold}, G. and {Gelly}, S. and {Uszkoreit}, J. and {Houlsby}, N.},
       title = "{An Image is Worth 16x16 Words: Transformers for Image Recognition at Scale}",
     journal = {arXiv e-prints},
         year = 2020,
        month = oct,
          eid = {arXiv:2010.11929},
        pages = {arXiv:2010.11929},
          doi = {10.48550/arXiv.2010.11929},
archivePrefix = {arXiv},
       eprint = {2010.11929},
 primaryClass = {cs.CV},
       adsurl = {https://ui.adsabs.harvard.edu/abs/2020arXiv201011929D}
}

@ARTICLE{cerezo2021,
      author = {{Cerezo}, M. and {Arrasmith}, J. and {Babbush}, R. and {Benjamin}, S.~C. and {Endo}, S. and {Fujii}, K. and {McClean}, J.~R. and {Mitarai}, K. and {Yuan}, X. and {Cincio}, L. and {Sornborger}, P.~J.},
       title = "{Variational quantum algorithms}",
     journal = {Nature Reviews Physics},
        year = 2021,
      volume = {3},
      number = {9},
       pages = {625--644},
          doi = {10.1038/s42254-021-00348-9},
archivePrefix = {arXiv},
       eprint = {2012.09265},
 primaryClass = {quant-ph},
       adsurl = {https://ui.adsabs.harvard.edu/abs/2021NatRP...3..625C}
}

@article{biamonte_quantum_2017,
	title = {Quantum machine learning},
	volume = {549},
	issn = {1476-4687},
	url = {https://doi.org/10.1038/nature23474},
	doi = {10.1038/nature23474},
	number = {7671},
	journal = {Nature},
	author = {Biamonte, Jacob and Wittek, Peter and Pancotti, Nicola and Rebentrost, Patrick and Wiebe, Nathan and Lloyd, Seth},
	month = sep,
	year = {2017},
	pages = {195--202},
}

@inproceedings{grover1996,
  author    = {Grover, Lov K.},
  title     = {A Fast Quantum Mechanical Algorithm for Database Search},
  booktitle = {Proceedings of the Twenty-Eighth Annual ACM Symposium on Theory of Computing},
  pages     = {212--219},
  year      = {1996},
  doi       = {10.1145/237814.237866}
}

@article{montanaro2016,
  author  = {Montanaro, Ashley},
  title   = {Quantum Algorithms: An Overview},
  journal = {npj Quantum Information},
  volume  = {2},
  pages   = {15023},
  year    = {2016},
  doi     = {10.1038/npjqi.2015.23}
}

@inproceedings{gilyen2019,
  author    = {Gily{\'e}n, Andr{\'a}s and Su, Yuan and Low, Guang Hao and Wiebe, Nathan},
  title     = {Quantum Singular Value Transformation and Beyond: Exponential Improvements for Quantum Matrix Arithmetics},
  booktitle = {Proceedings of the 51st Annual ACM SIGACT Symposium on Theory of Computing},
  year      = {2019}
}

@article{harrow2009,
  title = {Quantum Algorithm for Linear Systems of Equations},
  author = {Harrow, Aram W. and Hassidim, Avinatan and Lloyd, Seth},
  journal = {Phys. Rev. Lett.},
  volume = {103},
  issue = {15},
  pages = {150502},
  numpages = {4},
  year = {2009},
  month = {Oct},
  publisher = {American Physical Society},
  doi = {10.1103/PhysRevLett.103.150502},
  url = {https://link.aps.org/doi/10.1103/PhysRevLett.103.150502}
}

@inproceedings{yu_non-asymptotic_2024,
	title = {Non-asymptotic {Approximation} {Error} {Bounds} of {Parameterized} {Quantum} {Circuits}},
	volume = {37},
	url = {https://proceedings.neurips.cc/paper_files/paper/2024/file/b360fd42ced877429882a2a68b4a4343-Paper-Conference.pdf},
	doi = {10.52202/079017-3143},
	booktitle = {Advances in {Neural} {Information} {Processing} {Systems}},
	publisher = {Curran Associates, Inc.},
	author = {Yu, Zhan and Chen, Qiuhao and Jiao, Yuling and Li, Yinan and Lu, Xiliang and Wang, Xin and Yang, Jerry Zhijian},
	editor = {Globerson, A. and Mackey, L. and Belgrave, D. and Fan, A. and Paquet, U. and Tomczak, J. and Zhang, C.},
	year = {2024},
	pages = {99089--99127},
}

@book{du_gentle_2025,
	address = {Singapore},
	title = {A {Gentle} {Introduction} to {Quantum} {Machine} {Learning}},
	copyright = {https://www.springernature.com/gp/researchers/text-and-data-mining},
	isbn = {9789819512836 9789819512843},
	url = {https://link.springer.com/10.1007/978-981-95-1284-3},
	doi = {10.1007/978-981-95-1284-3},
	language = {en},
	urldate = {2026-09-09},
	publisher = {Springer Nature},
	author = {Du, Yuxuan and Wang, Xinbiao and Guo, Naixu and Yu, Zhan and Qian, Yang and Zhang, Kaining and Hsieh, Min-Hsiu and Rebentrost, Patrick and Tao, Dacheng},
	year = {2025},
}

@article{schuld2019qmlFeatureHilbertSpace,
  title = {Quantum Machine Learning in Feature Hilbert Spaces},
  author = {Schuld, Maria and Killoran, Nathan},
  journal = {Phys. Rev. Lett.},
  volume = {122},
  issue = {4},
  pages = {040504},
  numpages = {6},
  year = {2019},
  month = {Feb},
  publisher = {American Physical Society},
  doi = {10.1103/PhysRevLett.122.040504},
  url = {https://link.aps.org/doi/10.1103/PhysRevLett.122.040504}
}

@article{cerezo_challenges_2022,
	title = {Challenges and opportunities in quantum machine learning},
	volume = {2},
	issn = {2662-8457},
	url = {https://doi.org/10.1038/s43588-022-00311-3},
	doi = {10.1038/s43588-022-00311-3},
	number = {9},
	journal = {Nature Computational Science},
	author = {Cerezo, M. and Verdon, Guillaume and Huang, Hsin-Yuan and Cincio, Lukasz and Coles, Patrick J.},
	month = sep,
	year = {2022},
	pages = {567--576},
}

@article{huang_power_2021,
	title = {Power of data in quantum machine learning},
	volume = {12},
	issn = {2041-1723},
	url = {https://doi.org/10.1038/s41467-021-22539-9},
	doi = {10.1038/s41467-021-22539-9},
	number = {1},
	journal = {Nature Communications},
	author = {Huang, Hsin-Yuan and Broughton, Michael and Mohseni, Masoud and Babbush, Ryan and Boixo, Sergio and Neven, Hartmut and McClean, Jarrod R.},
	month = may,
	year = {2021},
	pages = {2631},
}

@misc{zhang_bits_2026,
	title = {From {Bits} to {Qubits}: {The} {Theory} and {Practice} of {Quantum} {Data} {Encoding}},
	shorttitle = {From {Bits} to {Qubits}},
	url = {http://arxiv.org/abs/2609.08058},
	doi = {10.48550/arXiv.2609.08058},
	urldate = {2026-09-09},
	publisher = {arXiv},
	author = {Zhang, Xiao-Ming and Rattew, Arthur G. and Wu, Bujiao and Styliaris, Georgios and Sun, Xiaoming and Koczor, Bálint and Yuan, Xiao},
	month = sep,
	year = {2026},
	note = {arXiv:2609.08058 [quant-ph]},
}

@article{aaronson2015fineprint,
  author  = {Aaronson, Scott},
  title   = {Read the fine print},
  journal = {Nature Physics},
  volume  = {11},
  pages   = {291--293},
  year    = {2015},
  doi     = {10.1038/nphys3272}
}

@article{fowler2012surface,
  author  = {Fowler, Austin G. and Mariantoni, Matteo and
             Martinis, John M. and Cleland, Andrew N.},
  title   = {Surface codes: Towards practical large-scale quantum computation},
  journal = {Physical Review A},
  volume  = {86},
  pages   = {032324},
  year    = {2012},
  doi     = {10.1103/PhysRevA.86.032324}
}

@article{hoefler2023disentangling,
  author  = {Hoefler, Torsten and H{\"a}ner, Thomas and Troyer, Matthias},
  title   = {Disentangling Hype from Practicality:
             On Realistically Achieving Quantum Advantage},
  journal = {Communications of the ACM},
  volume  = {66},
  number  = {5},
  pages   = {82--87},
  year    = {2023},
  doi     = {10.1145/3571725}
}
\clearpage

\begin{table*}
\centering
\caption{Performance of the selected ViT and QViT configurations on the main test set and the source-held-out FRB~20220912A set. Each configuration was selected by the highest mean validation recall within its model family, using common search ranges for corresponding architectural parameters. Values are mean $\pm$ standard deviation over 10 independent seeds.}
\label{tab:main_performance}
\begin{tabular}{lccccc}
\toprule
Model  &  Dataset & Accuracy (\%) & Recall (\%) & Precision (\%) & $F_2$ (\%) \\
\midrule
ViT & test & $93.83\pm0.60$ & $98.23\pm0.22$ & $90.30\pm1.02$ & $96.54\pm0.26$ \\
QViT &  test & $94.00\pm0.48$ & $98.30\pm0.60$ & $90.52\pm0.62$ & $96.64\pm0.50$ \\
ViT &  FRB 20220912A & $95.09\pm0.54$ & $97.90\pm0.74$ & $92.69\pm0.51$ & $96.80\pm0.64$ \\
QViT &  FRB 20220912A & $94.91\pm0.72$ & $97.11\pm1.31$ & $93.07\pm1.81$ & $96.26\pm0.82$ \\
\bottomrule
\end{tabular}
\end{table*}

\begin{table*}
\centering
\caption{Classification outcomes of QViT inference on the final balanced hardware-validation subset containing 41 FRB-like and 41 RFI segments.}
\label{tab:real_confusion}
\begin{tabular}{lcccc}
\toprule
Model & FRB-like Correct & FRB-like Incorrect & RFI Correct & RFI Incorrect \\
\midrule
QViT (simulator) & 33 & 8 & 40 & 1 \\
QViT (real device) & 34 & 7 & 34 & 7 \\
\bottomrule
\end{tabular}
\end{table*}


\begin{figure*}
\centering
\includegraphics[width=\linewidth]{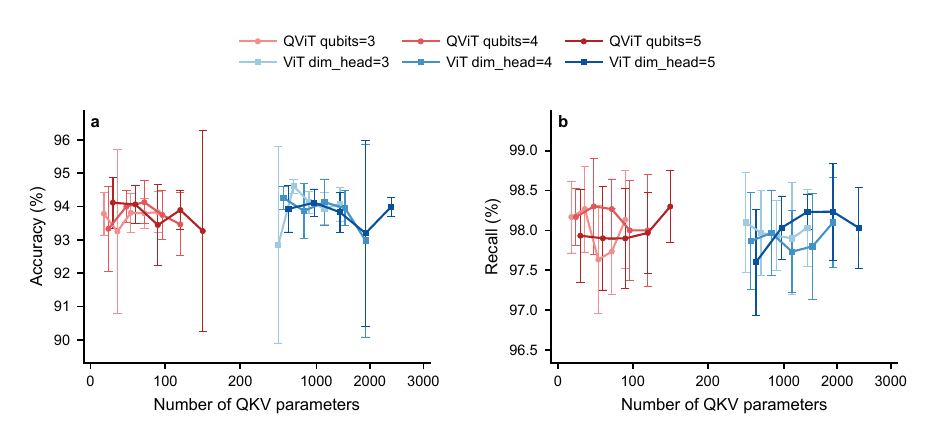}
\caption{
Performance comparison between QViT and classical ViT on the untouched test set. The horizontal axis denotes the number of trainable parameters in the Q/K/V projection module, computed as the total Q/K/V parameter count for each plotted configuration. (a) Classification accuracy. (b) Recall. Symbols indicate the mean over 10 independent seeds and error bars represent one standard deviation. For the selected comparison used in the main text, QViT reaches an accuracy of $94.00\pm0.48\%$ and a recall of $98.30\pm0.60\%$, compared with $93.83\pm0.60\%$ accuracy and $98.23\pm0.22\%$ recall for ViT.
}
\label{fig_res_compare}
\end{figure*}

\clearpage
\subsubsection*{Data availability}
Raw PSRFITS data are available from the FAST data center, \url{http://fast.bao.ac.cn}. The processed dynamic-spectrum segments, class labels are available are available at \url{https://github.com/euli0323/qvit-frb-code/tree/main}.

\subsubsection*{Code availability}
The dynamic spectrum plotting and QViT analysis code are available at \url{https://github.com/euli0323/qvit-frb-code/tree/main}.

\subsubsection*{Acknowledgements}
We acknowledge the use of artificial intelligence tools for language editing and code optimization during the preparation of this manuscript. This work is supported by the National Natural Science Foundation of China (grant No.\ 12503058, 12573096, 12603086), the Postdoctoral Fellowship Program of CPSF (grant No.\ GZC20252100), the China Postdoctoral Science Foundation (grant No.\ 2025M773201), the Jiangsu Funding Program for Excellent Postdoctoral Talent, the National Key Research and Development Plan of the Ministry of Science and Technology of China (Grant No. 2025YFE0212600), and the Science and Technology Program of Hebei Province of China (Grant No. 26391701D). A.R.Q. and C.F. thank the Paraíba State Research Foundation (FAPESQ-PB) and the Government of the State of Paraiba, Brazil.  The work by C. F. is supported by the CNPq (project PQ Grant 1A No. 311781/2021-7).  A.R.Q. acknowledges the financial support by CNPq under process number 306884/2026-7.  

\subsubsection*{Author Contributions}
L.H., X.Y. and S.B.Z. led data preparation and processing. S.B.Z., X.P.D. and X.F.W. initiated the hardware--algorithm co-design and coordinated regular project discussions. X.Y. and S.B.Z. developed the data processing pipeline prior to quantum integration, while L.H., Q.H.C., F.X. and A.R.Q. were responsible for the code quantization details and validation on the physical quantum hardware. All authors contributed to data analysis or interpretation and to the final version of the manuscript.

\subsubsection*{Competing Interests}
The authors declare that they have no competing financial interests.

\subsubsection*{Correspondence}
Correspondence and requests for materials should be addressed to S.B Zhang, Q.H. Chen, A.R. Queiroz, X.P. Deng or X.F. Wu.
\clearpage


\setcounter{table}{0} 
\captionsetup[table]{name={\bf Extended Data Table}}

\begin{table*}
\centering
\caption{Composition of the training, validation and evaluation sets. \textbf{(a)} Astrophysical pulses used to construct the FRB-like class, grouped by training source. \textbf{(b)} Segments in each data partition. The negative class consists of segments from single-pulse candidates reported by the same search pipeline that manual vetting rejected as RFI or other non-astrophysical signals. FRB 20220912A is excluded from training and from model selection and is used only for source-held-out evaluation. All images related to the same physical pulse, neighbouring raw-data interval or observing unit are assigned to a single partition.}
\label{tab:datasets}
\begin{tabular}{llc}
\toprule
\multicolumn{3}{l}{\textbf{(a) FRB-like class by training source}} \\
\midrule
Source & Type & Astrophysical pulses \\
\midrule
FRB 20220529 & Repeating FRB & 1243 \\
FRB 20201124A & Repeating FRB & 1093 \\
RRAT J1913+1330 & RRAT & 664 \\
\midrule
Total & & 3000 \\
\bottomrule
\end{tabular}

\vspace{1.5em}

\begin{tabular}{lccc}
\toprule
\multicolumn{4}{l}{\textbf{(b) Data partitions}} \\
\midrule
Partition & FRB-like & RFI & Total \\
\midrule
Training & 2400 & 2400 & 4800 \\
Validation & 300 & 300 & 600 \\
Test & 300 & 300 & 600 \\
\midrule
Total & 3000 & 3000 & 6000 \\
\midrule
FRB 20220912A (source-held-out) & 114 & 114 & 228 \\
\bottomrule
\end{tabular}
\end{table*}

\begin{table*}
\centering
\caption{Class composition of the test set after stratification by $\mathrm{SNR}=12$. \textbf{(a)} Number of FRB-like and RFI segments in each SNR subset. \textbf{(b)} Recall (per cent) of the selected ViT and QViT configurations, reported as mean $\pm$ standard deviation over 10 independent seeds. The hardware-validation subset is drawn from the low-SNR subset; because that subset contains 68 FRB-like segments, a difference of one segment corresponds to 1.47 percentage points of recall.}
\label{tab:snr}
\begin{tabular}{lccc}
\toprule
\multicolumn{4}{l}{\textbf{(a) Class composition}} \\
\midrule
Subset & FRB-like & RFI & Total \\
\midrule
Low SNR ($\mathrm{SNR}<12$) & 68 & 163 & 231 \\
High SNR ($\mathrm{SNR}\geq12$) & 232 & 137 & 369 \\
\midrule
Total & 300 & 300 & 600 \\
\bottomrule
\end{tabular}

\vspace{1.5em}

\begin{tabular}{lcc}
\toprule
\multicolumn{3}{l}{\textbf{(b) Recall (per cent)}} \\
\midrule
Model & Low SNR ($\mathrm{SNR}<12$) & High SNR ($\mathrm{SNR}\geq12$) \\
\midrule
ViT & $95.88\pm0.93$ & $98.92\pm0.23$ \\
QViT & $96.62\pm1.39$ & $98.79\pm0.53$ \\
\bottomrule
\end{tabular}
\end{table*}

\setcounter{figure}{0} 
\captionsetup[figure]{name={\bf Extended Data Figure}}

\begin{figure*}
\centering
\includegraphics[width=\linewidth]{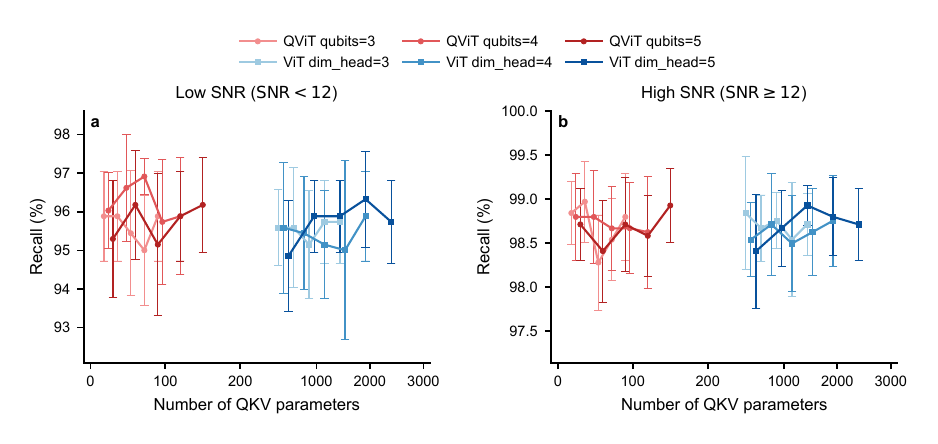}
\caption{
Recall comparison between QViT and classical ViT after stratifying the test set by signal-to-noise ratio. (a) Low-SNR subset, defined as $\mathrm{SNR}<12$. (b) High-SNR subset, defined as $\mathrm{SNR}\geq12$. The horizontal axis denotes the number of trainable parameters in the Q/K/V projection module. Symbols indicate the mean recall over 10 independent seeds and error bars represent one standard deviation. The SNR threshold is fixed independently of model performance and is applied to both FRB-like positive and negative segments using the SNR value encoded in each segment file name. For the selected comparison used in the main text, QViT reaches a low-SNR recall of $96.62\pm1.39\%$, compared with $95.88\pm0.93\%$ for ViT. The corresponding high-SNR recalls are $98.79\pm0.53\%$ and $98.92\pm0.23\%$, respectively.}
\label{fig_recall_snr_compare}
\end{figure*}

\begin{figure*}
\centering
\includegraphics[width=\linewidth]{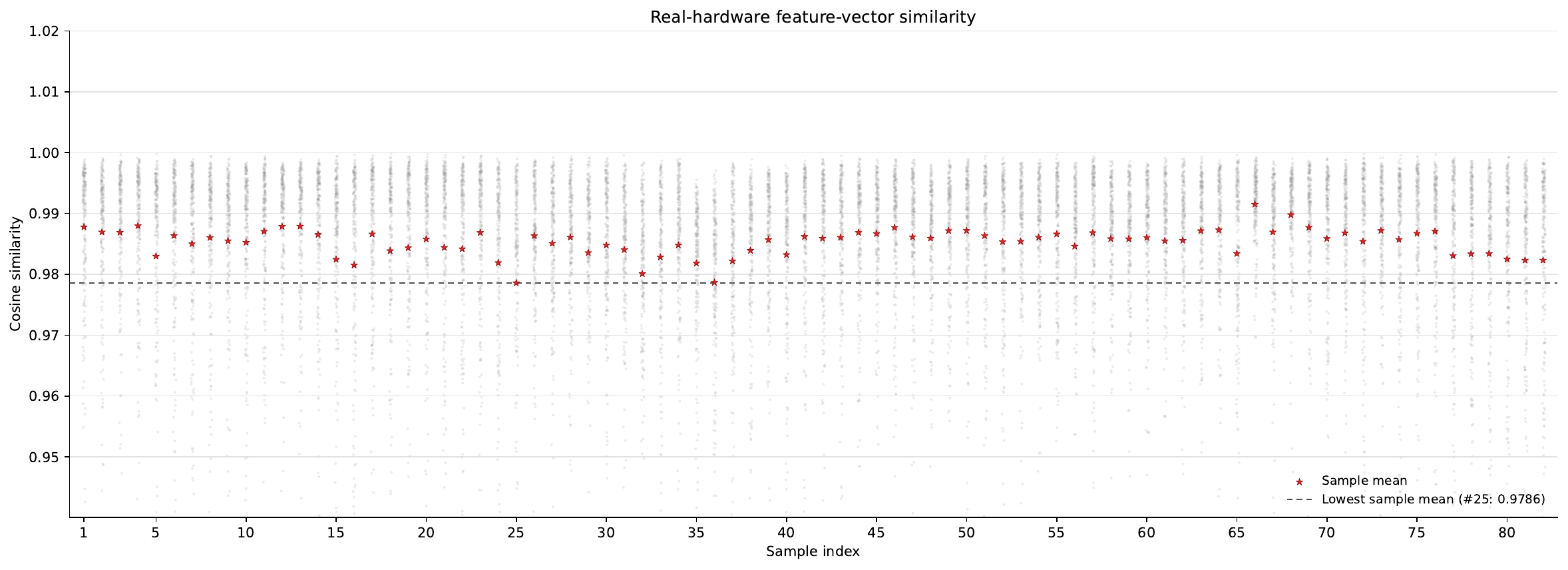}
\caption{Cosine similarity between simulated and real-device quantum feature vectors for the real-hardware validation samples. For each segment, the cosine similarity is averaged over the expectation-value comparisons from its 204 Q/K/V circuit evaluations. Across 82 validation segments, the minimum segment-level mean similarity is 0.9786.}
\label{fig_real_cosine}
\end{figure*}

\end{document}